# The Rise of the Smart Compound: Privately Governed Urban Intelligence and Its Research Agenda

Stefaan G. Verhulst

**Abstract**

*Across a range of fast-growing urban markets, private developers are constructing a version of the smart city that operates largely outside the purview of municipal government — often at the explicit invitation of city officials seeking to shift the cost and complexity of digital infrastructure onto private capital. Gated residential and mixed-use developments are increasingly marketed not merely on the basis of security and amenity, but on their "smartness": integrated home automation, app-mediated access control, and centralized energy and resource management, among other features. We refer to this phenomenon as the "smart compound." Despite its rapid proliferation, it has received comparatively little sustained scholarly attention: the literatures on smart cities and on gated communities have developed largely independently of one another, even as developers are, in practice, merging the two. This paper introduces the smart compound as an emerging real estate and urban development phenomenon, considers the opportunities and risks it presents, and examines how questions of data governance differ when smart urban infrastructure is built and owned privately rather than publicly. It concludes with a set of research questions intended to orient researchers, planners, and regulators toward a phenomenon whose growth is outpacing the scholarship meant to account for it.*

## 1. Introduction: A Growing Unit of Urban Intelligence

For most of the past two decades, the "smart city" has been the dominant frame for thinking about technology and urban life: municipal governments, often working with large technology vendors, layering sensors, data platforms, and digital services across an entire metropolitan area. But a quieter and arguably faster-moving phenomenon has been building alongside it. Private real estate developers, not city governments, are now building "smart" at a much smaller scale: the gated residential compound, the master-planned community, the private mixed-use district.

The term "smart compound" is not yet an established academic category, but it is already a marketing and product category in fast-growing real estate markets. In Egypt, for instance, developers building gated communities around Cairo now routinely brand entire projects as smart compounds — integrating home automation, app-based access control, centralized security, and building-level energy management as standard features rather than add-ons; in projects such as Nyoum and iCity New Cairo, or Nha Be in Ho Chi Minh City, spanning hundreds of acres and thousands of units. The pattern is not unique to Egypt or Vietnam: it echoes across Gulf megaprojects, Chinese gated new towns, and master-planned suburbs in Southeast Asia and Latin America - wherever private capital is building large residential or mixed-use developments from scratch.

What makes this a distinct phenomenon, rather than simply "smart city, but smaller", is who is doing the building, and under what structure of control and accountability. It is tempting to draw the line as public versus private: smart cities as public projects — initiated by municipalities, financed (at least partly) by public budgets, and accountable, however imperfectly, to residents and elected officials — and smart compounds as private products, initiated by developers, financed by land and unit sales, and accountable primarily to buyers and investors (and to public officials mainly at the permitting stage). But the line is better drawn as a continuum than a wall. Some of the most prominent compound-scale developments are instruments of public strategy: Singapore's Punggol Digital District is gated-adjacent, sensor-saturated, and run on a single integrated platform, yet was built by a state agency as a vehicle of the national Smart Nation strategy; Korea has distributed a national smart-district platform to more than a hundred local governments; and joint public–private "one district" models are an explicit rollout strategy elsewhere. Conversely, municipal smart cities increasingly run on privately supplied platforms and terms (Microsoft in Seoul; Waymo in Tokyo). What varies along this continuum is who owns the land and infrastructure, who controls the data, through what channel residents can hold the operator to account, and how autonomous the development's digital systems are. It is the far end of that continuum that concerns us here: the fully private smart compound, where "smartness" is a feature engineered to raise property values and differentiate a project in a crowded market, not a public good delivered through municipal policy. That position, and the incentive structure that comes with it, turns out to matter enormously for how these places function, who benefits from them, and how connected they are to the cities around them.

## 2. The Research Gap: Two Literatures That Rarely Meet

Academic research on smart cities is, by now, vast. Systematic reviews of the field have catalogued hundreds of publications spanning governance, sustainability, citizen participation, and technology architecture. But that same body of research consistently flags its own limitations: reviewers analyzing the smart city literature have found that research in the field is often fragmented and heavily technology-driven, skews toward documenting perceived benefits rather than downsides or failures, and still lacks strong theoretical grounding or empirical testing of its own frameworks.

Running in parallel, and largely separate, is a much older research tradition on gated communities, dating to the 1990s, that has produced its own systematic reviews spanning safety, social cohesion, exclusion, and urban fragmentation. Recent reviews in this space explicitly note that surveillance and AI technologies are now reshaping gated communities, but flag that the ethical and privacy implications of this technology layer remain under-studied, and call for more research into the balance between security gains and social costs. A related strand of scholarship on "residential compounds" has argued that researchers need a broader, more flexible concept than the classic "gated community" to capture the full diversity of these developments, since strict physical enclosure is no longer even a defining feature in many markets.

The gap is that these two literatures (smart cities and gated/private compounds) have developed almost independently. Smart city scholarship largely assumes a public-sector actor and a citywide scale, and treats fragmentation as something that happens *within* smart city governance (across agencies, vendors, and data systems) rather than as something that happens *between* a privately governed enclave and the public city around it. Gated-community scholarship, meanwhile, has extensively documented social and spatial fragmentation but has only recently begun to reckon with what happens when those enclosed developments also become data-collecting, sensor-equipped, algorithmically managed environments. Very little existing research asks the two questions together: what changes about the classic critique of gated communities when the walls come with cameras, apps, and predictive systems attached; and what changes about the smart city model when its most technologically dense environments are privately owned, privately governed, and legally separate from the city's own systems?

This is the analytical space that "smart compounds" occupies; and it is currently more of a live phenomenon on the ground than a settled field of study.

## 3. Opportunities: Compounds as Laboratories

Despite the risks discussed below, the smart-compound model creates some genuine advantages that public smart city programs often struggle to replicate.

### 3.1 Speed and coherence of implementation

Public smart city initiatives are frequently slowed by fragmented authority: multiple agencies, overlapping mandates, and the absence of a single body empowered to set and enforce a unified technology strategy. Research on smart city governance in developing countries has identified this fragmented authority (the lack of a central coordinating body and the resulting duplication and misalignment across agencies) as one of the most persistent barriers to effective rollout. A private developer building a single compound faces less of that coordination problem: it owns the land, develops the master plan, and can specify a single integrated technology stack from day one, rather than retrofitting legacy infrastructure or negotiating across departments.

### 3.2 Testbed function

Much of the innovation-district literature treats purpose-built, privately or jointly developed districts as testbeds where new technologies can be piloted at a contained scale before wider deployment — robotics and renewable energy technology trialed at Masdar City in the UAE, or autonomous vehicle systems piloted in Singapore's Jurong Innovation District, are cited as examples of this pattern. Smart compounds can serve an analogous function for residential and consumer technology: home automation, app-based access and payments, and localized energy or water management systems can be proven out in a single, contained development before assumptions about resident adoption and technical reliability are tested at city scale.

### 3.3 Hubs of innovation and talent clustering

The broader innovation-district literature, covering mixed-use districts such as Boston's Seaport District, Zürich-West, and Kendall Square, describes how master-planned, technology-dense environments can concentrate startups, research institutions, and skilled workers in ways that generate real economic spillovers, with one case study crediting a converted industrial district with attracting hundreds of new businesses and thousands of jobs over several years. Smart compounds that incorporate coworking space, business incubation, or anchor employers alongside housing are, in effect, applying that innovation-district logic to primarily residential products, positioning themselves as environments that attract higher-income, technology-oriented residents and, by extension, service and retail investment.

### 3.4 Governance agility

Because a compound operates under a single ownership or homeowners'-association structure rather than municipal bureaucracy, decisions about upgrading infrastructure, adjusting service pricing, or rolling out new digital services can be made and implemented far faster than in a public system that requires budget cycles, procurement processes, and political buy-in (see also table below).

## 4. Challenges: Fragmentation and Disconnection

The same features that make smart compounds fast and coherent to build are what generate their most serious downsides, and the concerns here are not hypothetical; they extend directly out of decades of gated-community research.

### 4.1 Spatial and social fragmentation

A substantial body of research on gated developments describes contemporary urban space as increasingly fragmented into unconnected enclaves, walled off both physically and socially from their surroundings. A study of gated neighborhoods in Shenzhen found that this enclosure pattern produces measurable disparities in residents' access to and satisfaction with amenities between gated and non-gated areas, and argued that the broader consequence is an increase in societal fragmentation driven by the privatization of what used to be shared urban space. Layering smart technology onto that structure does not resolve the fragmentation; if anything, it can reinforce it by giving the enclave its own self-contained digital ecosystem — its own app, its own data platform, its own service layer — that has no reason to interoperate with the city's public systems.

### 4.2 Disconnection from municipal governance and infrastructure

Smart compounds are typically designed as self-sufficient enclosures: private security substituting for public policing, private utilities or metering substituting for municipal systems, private data platforms substituting for any citywide digital infrastructure. Research on gated communities in Cairo has argued that this pattern of enclosure can gradually transform the social fabric of a city as more affluent, higher-status residents opt out of shared urban life and public services altogether. From a technology standpoint, this means the most sophisticated, sensor-rich, data-generating environments in a metropolitan area may sit entirely outside any municipal smart city strategy; their data unshared, their systems non-interoperable, their governance accountable to a developer or homeowners' association rather than to the city.

### 4.3 Exclusion and inequitable access to "smart" living

Because smart compounds are market products, access to their benefits is gated by price. Research on smart city development already warns of the risk that such initiatives can reinforce urban inequality and adopt a technocratic approach that sidelines the needs of lower-income residents; a privately built and privately priced smart compound makes that exclusion explicit and structural rather than an unintended side effect of public policy. The "smart city," whatever its flaws, is nominally offered to an entire municipal population; the smart compound is, by design, offered only to those who can buy in.

### 4.4 Ambiguous safety and autonomy trade-offs

Systematic reviews of gated communities have also surfaced more granular social costs that carry over directly into the smart-compound context — for example, evidence that increased surveillance and enclosure can create a paradox for some residents, particularly women, who report feeling both safer and more confined within these developments — a tension the reviewers argue needs much more dedicated research as surveillance technology becomes more pervasive.

### 4.5 Interoperability and standardization problems

Even setting aside governance questions, structured literature reviews of smart-city technology point out that systems built by different vendors for different developments frequently fail to interoperate, creating fragmentation and compatibility problems at the technical level. A city with dozens of independently developed smart compounds, each running its own vendor's proprietary platform, risks ending up with a patchwork of disconnected smart micro-environments rather than anything resembling a coherent citywide system — a technical mirror of the social fragmentation described above.

## 5. Data Governance: Smart Cities versus Smart Compounds

Municipal smart city programs and smart compounds are often discussed as though they raise the same data governance questions at different scales. In practice, the underlying governance logic is different enough that the same terms such as "data governance," "privacy," "accountability", can mean fairly different things in each setting. Table 1 summarizes the contrast across eight dimensions.

**Table 1.** Data governance in municipal smart cities versus privately governed smart compounds.

| Dimension | Smart cities | Smart compounds |
|---|---|---|
| **Legal basis for collecting data** | Public authority, typically shaped by national/regional data protection law, procurement rules, and (in some places) open-data legislation — e.g., the EU's open-data and non-personal-data-flow directives that European cities operate under. | (Mostly) Private contract — embedded in the sale agreement, HOA bylaws, or app terms of service a resident accepts when buying or leasing a unit; rarely shaped by data-specific legislation designed for this context. |

| Dimension | Smart cities | Smart compounds |
|---|---|---|
| **Who "owns" the data** | Contested but at least nominally a public question — cities are increasingly expected to define ownership standards for citizen-generated data before deciding what to publish, share, or restrict. | Typically the developer or its technology vendor by default; ownership terms are set unilaterally and rarely negotiated by individual residents. |
| **Oversight and accountability mechanism** | At least formally accountable through elected officials, public records law, and (in leading examples) dedicated privacy and surveillance oversight functions, such as Portland's Smart City PDX program, which pairs an open-data pillar with a separate privacy and surveillance-oversight pillar. | Accountable to a developer's board or an HOA, which residents did not elect and may have limited power to change; no external public oversight body reviews compound-level data practices. |
| **Primary purpose of data collection** | Framed (however imperfectly in practice) around a public good — traffic management, energy efficiency, emergency response — with data reuse for the "common good" treated as a policy goal in leading frameworks. | Framed around property value, security, and service delivery to paying residents; any public-good value (e.g., traffic or environmental data useful to the city) is incidental and not contractually guaranteed. |
| **Transparency and disclosure norms** | A recognizable model exists: open-data portals, published policies, impact assessments — imperfectly and unevenly applied, but with visible best-practice examples (Barcelona, Amsterdam, Bristol) that researchers and journalists can benchmark against. | No equivalent norm. Disclosure is usually a clause inside marketing material or a sales contract rather than a distinct data-governance document, and there is no established benchmark of "good practice" to hold a compound to. |
| **Interoperability with other systems** | An explicit policy goal in leading frameworks — cities pursue standardized formats and data-sharing agreements (e.g., Barcelona's procurement clauses requiring vendors to hand over public-value data) precisely because fragmentation across agencies is recognized as a problem. | Rarely a goal at all. Vendor lock-in to a single proprietary platform is often the default, and there is generally no contractual or regulatory expectation that a compound will share data with the municipality it sits inside. |
| **Resident/citizen recourse** | Imperfect but existing: public records requests, elected representatives, regulatory complaints, media scrutiny. | Minimal: recourse is generally limited to whatever complaint or dispute-resolution mechanism the developer's or HOA's contract provides, with no independent regulator typically assigned to review it. |
| **Dominant failure mode identified in the literature** | Fragmented authority *within* the public sector — overlapping agencies, inconsistent standards, and uneven enforcement of otherwise-reasonable governance principles. | Fragmentation *between* the private enclave and the public city: a well-run, internally coherent system that simply has no obligation or channel to connect to municipal data governance at all. |

In sum: smart city data governance research is mostly about fixing a public system that has the *right structure* but struggles to coordinate it. Smart compound data governance is closer to an open question of whether the right structure exists at all, which is precisely why applied questions such as model data-sharing agreements, regulatory gap-mapping, and consent-disclosure design (Section 7) are where near-term progress is most possible.

## 6. Bridging the Two Research Traditions

Closing the gap between smart-city research and gated/compound research would likely require a few specific shifts in how the phenomenon is studied:

- **Treating compounds as smart city infrastructure, not just housing.** Much smart city research still implicitly treats the "smart" layer as something municipalities build and own. Compounds that generate comparable — or greater — volumes of sensor and behavioral data, but sit under private ownership, arguably need to be brought inside the same analytical frame, including questions about data governance, interoperability, and public accountability that smart city researchers already ask of municipal systems.
- **Bringing technology and surveillance questions into gated-community research at the same depth as social and spatial ones.** Gated-community scholars have spent thirty years building sophisticated tools for studying social fragmentation, exclusion, and safety perception. Applying that same rigor to the specific technologies now embedded in these developments — rather than treating "smart" as a marketing footnote — would let researchers ask sharper questions about consent, data ownership, and algorithmic management inside private enclosures.
- **Comparative, cross-market case studies.** Because "smart compound" branding has emerged somewhat independently in different regions — Egyptian and Gulf real estate markets, Chinese gated new towns, Latin American master-planned communities — there is an open opportunity for comparative research asking whether the opportunities and risks described above play out consistently across very different regulatory, cultural, and urban planning contexts, or whether local governance capacity changes the calculus significantly.
- **Longitudinal study of the disconnect problem.** Almost all of the fragmentation research cited above is cross-sectional — a snapshot of a compound or gated development at one point in time. Given how quickly the smart-compound product category is expanding, longitudinal research tracking whether these developments become more or less integrated with surrounding municipal systems over time would be especially valuable for planners deciding how to respond.

## 7. A Preliminary Research Agenda

The questions below are organized by theme and numbered consecutively (RQ1–RQ26). They are intended as entry points for researchers, planners, and regulators rather than as an exhaustive catalogue.

### 7.1 Conceptual and definitional

**RQ1.** What distinguishes a "smart compound" from a smart city district, an innovation district, and a conventional gated community; and does the field need a formal typology, the way gated-community scholars have proposed a broader "residential compound" category to replace the narrower "gated community" label?

**RQ2.** At what scale (units, acreage, population) does a private development start to function like a parallel city rather than a neighborhood; and does that threshold matter for how it should be governed or regulated?

**RQ3.** Is "smart" in this context primarily a technology bundle, a marketing signal, or a genuine shift in how a development is planned and operated; and how would a researcher tell these apart empirically?

### 7.2 Fragmentation and urban integration

**RQ4.** Do smart compounds increase, decrease, or simply relocate the fragmentation already documented in gated-community research, once digital infrastructure (apps, sensors, private data platforms) is layered on top of physical walls?

**RQ5.** How do residents of smart compounds actually use, or avoid using, public city services and infrastructure once private, app-based alternatives exist inside the compound?

**RQ6.** What happens at the literal edge of a smart compound: how do adjacent, non-gated neighborhoods experience spillover effects (traffic, service demand, land values, visual and social separation) from a technologically dense enclave next door?

### 7.3 Data governance: conceptual questions

**RQ7.** Who controls the data generated inside a smart compound; and does "ownership" even capture the question? Given that trade secrecy and contract confer control rather than title, how are three layers allocated in practice — control over raw inputs, entitlement to derived insights and analytics, and access interfaces for residents, municipalities, and vendors — and how does that allocation vary across the markets where smart compounds are proliferating (Egypt, the Gulf, China, Latin America) and across regulatory regimes (e.g., the EU's Data Act, which grants users access to raw device data but leaves derived insights with the operator)?

**RQ8.** Should data generated inside a privately governed but densely populated compound be treated as private commercial data, or does its scale and public-interest relevance (traffic patterns, energy use, emergency response) argue for treating it more like public infrastructure data?

**RQ9.** What recourse do residents have if a developer sells, licenses, or discontinues the platform underlying the compound's "smart" systems (access control, security, utilities); and how does this compare to protections residents would have under a municipal system?

**RQ10.** How does data governance inside a smart compound interact with national data protection law, especially in markets where such law is new or unevenly enforced?

**RQ11.** Do compound-level surveillance and access-control systems (widely used for security marketing) create differential privacy burdens for staff, service workers, and visitors compared with resident-owners; and who consents to that data collection on their behalf?

### 7.4 Data governance: applied questions

**RQ12.** *Interoperability audit.* For a given city, what proportion of newly built smart compounds use proprietary, closed platforms versus open or standards-based systems; and what would it take and cost, technically and contractually, to make compound-level data (e.g., traffic, energy, water use) available to municipal planners in aggregated, privacy-preserving form?

**RQ13.** *Model data-sharing agreement.* What would a template data-sharing agreement between a private compound developer/HOA and a municipal government look like, specifying what categories of data are shared, at what level of aggregation, on what schedule, and under what privacy safeguards, and could such a template be piloted with a willing developer?

**RQ14.** *Comparative regulatory gap-mapping.* In a specific jurisdiction (e.g., Greater Cairo, a Gulf emirate, a Chinese municipality), which existing laws, data protection, telecommunications, urban planning, consumer protection, apply to compound-level smart systems, and where do clear gaps or contradictions exist that a regulator could act on?

**RQ15.** *Consent and disclosure design.* What would meaningfully informed consent look like for a resident purchasing a unit in a smart compound, given that today's disclosures are typically embedded in marketing materials and sales contracts rather than presented as data governance choices; and could a standardized disclosure format be tested with real buyers?

**RQ16.** *Emergency and public-safety access.* Under what conditions, and through what legal mechanism, should municipal emergency services or public health authorities be able to access compound-generated data (e.g., occupancy, access logs, environmental sensors) during a crisis, and how should that be balanced against the privacy commitments made to residents?

**RQ17.** *Vendor lock-in and continuity planning.* What contractual or regulatory safeguards would protect residents if the technology vendor supplying a compound's smart systems goes out of business, is acquired, or unilaterally changes its data practices; and do current developer contracts in fast-growing markets address this risk at all?

**RQ18.** *Comparative data-sharing pilot.* Would a matched-pair study (one smart compound with a voluntary municipal data-sharing arrangement and one without) show measurable differences in public service planning outcomes (traffic management, utility load balancing, emergency response times) in the surrounding area?

**RQ19.** *Liability allocation.* When a municipality plans or acts on compound-generated data that proves inaccurate, or a compound acts on faulty city data, who bears the loss, under what instrument, and how does the allocation of that risk shape each side's willingness to share at all?

### 7.5 Equity and social outcomes

**RQ20.** Does living in a smart compound change residents' expectations of, or willingness to pay for, public services elsewhere in the city; and does that have a measurable effect on political support for public smart city investment?

**RQ21.** How do smart compounds affect the workers who service them (domestic staff, security, maintenance) in terms of surveillance exposure, mobility within the compound, and data collected about their movements and work patterns?

**RQ22.** Does the "safety versus autonomy" paradox identified in gated-community research (particularly for women) intensify, ease, or simply change character when the security apparatus becomes app-based and data-driven rather than purely physical?

**RQ23.** Who pays for, and who profits from, compound intelligence? How do financing and value-capture arrangements (service subscriptions, data commercialization, property-value uplift) distribute the benefits of the "smart" layer among developer, residents, vendors, and the surrounding city, and do tiered or subscription-based access models create a digital divide within and around the compound?

### 7.6 Methodology

**RQ24.** Given that most existing evidence is cross-sectional, what would a longitudinal research design look like for tracking a smart compound's integration with (or continued separation from) its host city over a 10–15 year horizon?

**RQ25.** What comparative research design would allow researchers to isolate the effect of "smart" technology from the effect of "gated" enclosure; for instance, comparing gated-but-not-smart, smart-but-not-gated, and smart-and-gated developments within the same city?

**RQ26.** Most of the empirical questions above assume researchers can compare outcomes inside a smart compound with outcomes in the surrounding city, but compared on what? Existing smart-city index measures have not gone below the city level, and smart compounds currently report their performance in marketing terms of their own choosing. What shared indicator set, for instance, SDG indicators localized to district scale, as some cities already do in voluntary local reviews, would allow a smart compound's outcomes (energy use, safety, mobility, inclusion, resident satisfaction) to be measured on the same scale as a municipal program's, so that claims about the benefits and harms of smart compounds can be tested rather than asserted?

## 8. Conclusion

Smart compounds represent a genuine and under-examined shift in how urban technology gets built: not from the top down, through municipal smart city strategy, but from the ground up, through private developers competing for buyers in a crowded real estate market. That shift brings real advantages, such as faster implementation, coherent technology stacks, testbed potential, and innovation-district-style clustering effects. But it also imports, and in some ways sharpens, the long-standing critiques of gated urban development: fragmentation of urban space, disconnection from public governance and infrastructure, and exclusion structured directly by price.

It also connects to a conversation this paper has so far left implicit. The most developed compounds do not just deploy sensors; they run their own urban digital twins, at exactly the moment when cities are building twins of their own. How a privately owned, compound-scale twin relates to a municipal one is a special case of the federation problem the digital-twin literature is now working through: whether independently governed twins can share selectively, on trusted terms, without a single authority over the data. That literature also offers a caution against fatalism. Private twins have been built voluntarily on open standards, with contractual guarantees of responsible data use; private, in other words, need not mean ungoverned. The research question is why such alignment remains the exception rather than the rule.

The academic literature has, for decades, studied smart cities and gated communities as separate phenomena. As developers continue building compounds that are simultaneously gated and smart, closing that gap in the research is no longer optional — it is necessary to understand what kind of city is actually being built, one compound at a time.

## Acknowledgements

The author is grateful for input from Begoña Otero and Adam Zable on earlier versions of this paper.